\documentclass[pra,twocolumn,superscriptaddress]{revtex4-1}
\usepackage{graphicx}
\usepackage{amsmath,amssymb}
\usepackage{hyperref}
\usepackage{array}
\usepackage{placeins}

\begin{document}
\title{Entanglement Classification in the Graph States:
 The generalization to \textit{n} – Qubits States using the Entanglement Matrix} 
\author{Sameer Sharma}
\email{sameer20@iiserb.ac.in}
\affiliation{Indian Institute of Science Education and Research, Bhopal}
\begin{abstract}
Graph states represent a significant class of multi-partite entangled quantum states with applications in quantum error correction, quantum communication, and quantum computation. In this work, we introduce a novel formalism called the Entanglement Matrix for quantifying and classifying entanglement in n-qubit graph states. Leveraging concepts from graph theory and quantum information, we develop a systematic approach to analyze entanglement by identifying primary and secondary midpoints in graph representations, where midpoints correspond to controlled-Z gate operations between qubits. Using Von Neumann entropy as our measure, we derive precise mathematical relationships for maximum entanglement in graph states as a function of qubit number. Our analysis reveals that entanglement follows a quadratic relationship with the number of qubits, but with distinct behaviors for odd versus even qubit systems. For odd n-qubit graph states, maximum entanglement follows $E_{max} = n^2-n$, while even n-qubit states exhibit higher entanglement with varying formulae depending on specific configurations. Notably, systems with qubit counts that are multiples of 12 demonstrate enhanced entanglement properties. This comprehensive classification framework provides valuable insights into the structure of multi-qubit entanglement, establishing an analytical foundation for understanding entanglement distribution in complex quantum systems that may inform future quantum technologies.
\end{abstract}
\maketitle

\section{Introduction}
\noindent
Entanglement, a fundamental aspect of quantum mechanics, signifies the profound distinction between classical and quantum physics. It manifests as a variety of non-local quantum correlation subsystems, which have numerous applications in the quantum realm, such as quantum teleportation, quantum dense coding, quantum cryptography, and quantum computing \cite{Eisert2000Schmidt}\cite{Elsaman2023QuantumNetworks, Raussendorf2001OneWayModel}. Quantification and classification of quantum entanglement are required to determine the class of quantum states and the depth of the entanglement level, which impacts the performance of quantum information tasks \cite{Jozsa2006AnIntroduction}.

Graph states \cite{Wikipedia2023GraphStates} are a class of multi-partite entangled states that correspond to mathematical graphs, where the vertices of the graph take the role of quantum spin systems and edges represent Ising interactions \cite{Vesperini2023GraphConnectivity, Perlin2024Experimental}. They are many-body spin states of distributed quantum systems that play a significant role in quantum error correction, multi-party quantum communication, and quantum computation within the framework of the one-way quantum computer and the exploration of fundamental concepts like non-locality and decoherence \cite{Raussendorf2001OneWay, VanDenNest2004Clifford}. We characterize and quantify the genuine multi-particle entanglement of such graph states in terms of Schmidt measure \cite{Eisert2000Schmidt} and Von Neumann entropy measure \cite{Plenio2005Entanglement}, from where, we defined our entanglement matrix which can be used to give the relation between the maximum entanglement for any N Qubit graph state with respect to the number of qubits (N).

\section{Preliminaries}
\subsection*{Graph Theory and Analogy with Graph States}
In graph theory, we represent vertices by nodes. More formally, a graph \( G = (V, E) \) is a pair of sets where \( V \) contains a finite number of nodes, and \( E \subseteq V \times V \) is a multiset of edges connecting them \cite{Wikipedia2023GraphStates}. For undirected graphs, edges are unordered pairs of nodes. Two nodes connected by an edge are called adjacent nodes, and the degree of a node is the number of edges connected to that node. If a node has degree one, it is called an endpoint.

\subsubsection{Isomorphic Graphs}
Consider two simple graphs, \( G_1 = (V_1, E_1) \) and \( G_2 = (V_2, E_2) \). These graphs are isomorphic if there exists a bijective function \( f: V_1 \to V_2 \) such that two vertices \( v_i, v_j \) are adjacent in \( G_1 \) if and only if \( f(v_i) \) and \( f(v_j) \) are adjacent in \( G_2 \) \cite{Ryjacek1987}.

\subsubsection{Non-Isomorphic Graphs}
If two graphs are not isomorphic, they are considered non-isomorphic. This means that there is no bijection between their vertices that preserves the edge structure \cite{Babai2015}. Non-isomorphic graphs are distinct graphs that cannot be transformed into each other by a simple relabelling of vertices while preserving the edge connections.

Following the concepts from graph theory, in Quantum Information Theory, the nodes are replaced by 'qubits' and the edge connecting two nodes (if they exist) is replaced by a Controlled-Z gate known as the 'CZ gate'. This arrangement of qubits as nodes and CZ gates as edges leads us to another class of Quantum States known as Graph States.

\subsection*{Graph States}
One way to represent entangled states is through graph states using adjacency matrices. For a simple and undirected graph \( G = (V, E) \) of size \( N \) with adjacency matrix being a square matrix \( \Gamma_{N \times N} \) so that its \( \Gamma_{ij} = \Gamma_{ji} = 1 \) when there is an edge \( (i, j) \) in \( E \), and \( \Gamma_{ij} = \Gamma_{ji} = 0 \) when there is no edge. For every edge \( (i, j) \) present on the graph, we apply a \( CZ_{ij} \) Gate on the corresponding quantum circuit. The N-qubit graph state can be written as \cite{Hein2004,Hein2006GraphReview, Hein2006GraphStatesReview}:

\[
|G\rangle = \prod_{(i,j) \in E} CZ_{ij} |+\rangle^{\otimes N}
\]

Where, \( |+\rangle_x = \frac{1}{\sqrt{2}} (|0\rangle + |1\rangle) \) is an eigenstate of Pauli Operator \( \hat{\sigma}_x \), with eigenvalue +1.

Here are some examples of non-isomorphic classes for three and four-qubit graph states \cite{McKay1983}:
\begin{figure}[h]
    \centering
    \includegraphics[width=0.45\textwidth]{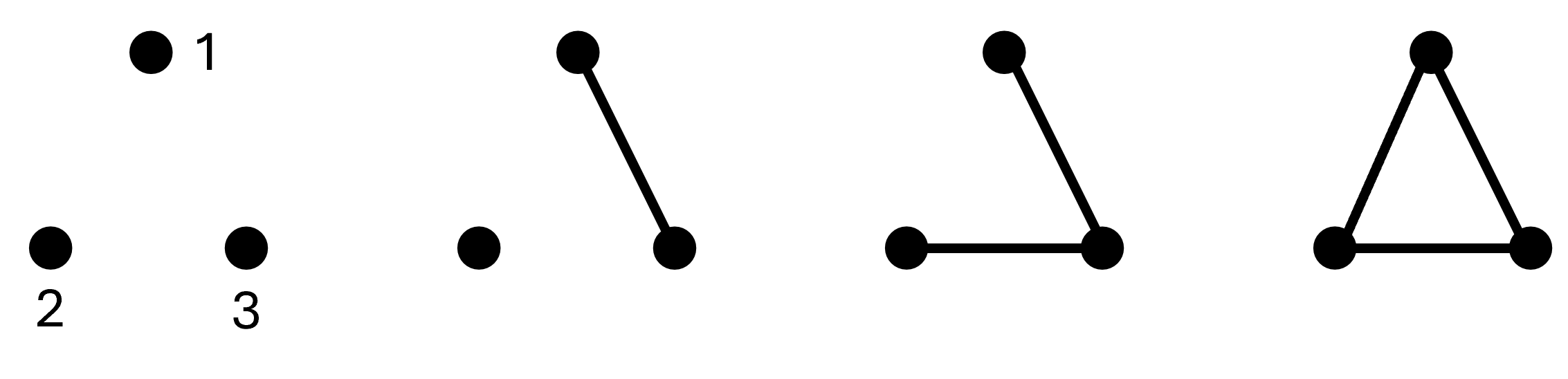}
    \caption{Four non-isomorphic classes for three-qubit graph states}
\end{figure}

\begin{figure}[h]
    \centering
    \includegraphics[width=0.47\textwidth]{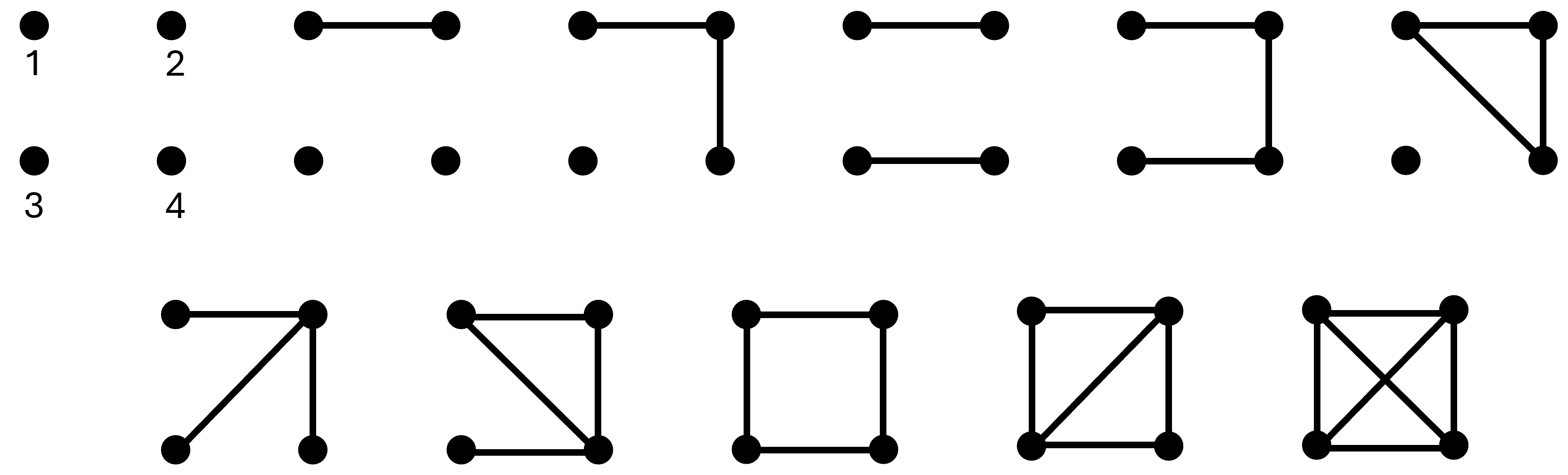}
    \caption{Eleven non-isomorphic classes for four qubit graph states}
\end{figure}

\subsection*{Entanglement in Graph States}

Understanding and characterizing entanglement in complex quantum systems like n-qubit graph states is crucial for leveraging their potential in quantum communication and computation tasks \cite{Dur2000Tripartite}. Some of the possible measures for classifying and quantifying the entanglement in non-isomorphic graph states are \cite{Guhene2005BellGraph, Guhne2005Verification}:

\subsection*{Graph Entropy Measures}

\subsubsection*{Von Neumann Entropy}

The Von Neumann Entropy quantifies the degree of mixing of the state describing a given finite system. It provides a measure of the uncertainty or randomness of the quantum state, with a lower entropy indicating a more certain or less random state. For pure states \( |\psi\rangle \) of a bipartite system, the entropy of formation is used as a measure of entanglement. The base of the logarithm, in this case, is base 2, and \( S \) is referred to as Von Neumann Entropy, which is \cite{vonNeumann1955}:

\begin{equation}
\begin{split}
    S(\rho_A) &= -Tr[\rho_A \log_2 \rho_A] = -Tr[\rho_B \log_2 \rho_B] \\
    &= - \sum_i p_i \log p_i
\end{split}
\end{equation}

Where A and B are the subsystems formed by partial trace of the original system of states and \( p_i \) are the corresponding eigenvalues of each subsystem (Both subsystems will have a common set of eigenvalues).

\subsubsection*{Rényi entropy}

For a density matrix \( \rho \in D(H) \), the quantum Rényi entropy is defined as follows \cite{Renyi1961}:

\[
S_\alpha(\rho) = \frac{1}{1-\alpha} \log Tr(\rho^\alpha), \alpha \in (0,1) \cup (1,\infty)
\]

This is a quantum version of a classical Rényi entropy. If \( \{p_i\}_i \) are the eigenvalues of \( \rho \), then the quantum Rényi entropy reduces to a Rényi entropy of a random variable \( X_\rho \) with probability distribution \( \{p_i\} \).

\[
S_\alpha(\rho) = H_\alpha(X_\rho) = \frac{1}{1 - \alpha} \log \left( \sum_i p_i^\alpha \right)
\]

As in the classical case, von Neumann entropy is a limiting case of the Rényi entropy, as \( \lim_{\alpha \to 1} S_\alpha(\rho) = S(\rho) \).\\

\textit{Tsallis Entropy} is another generalization of entropy that is used to quantify the entanglement. Similar to the Rényi entropy \cite{Tsallis1988}.

\subsection*{Graph Invariants}

Certain Graph invariants such as the Chromatic Number, Clique Number or girth can indirectly provide information about the entanglement properties. For Non-isomorphic graphs, these properties can differ significantly and lead to distinct entanglement characteristics \cite{VanDenNest2004Clifford, Schlingemann2001QEC, Markham2008Secret}.\\

In this paper, we will focus on entanglement classification based on Von-Neumann Entropy. In quantum information, the logarithms are usually taken to be base 2, giving a maximum entropy of 1 for a qubit.

\section{The Entanglement Matrix}

Before introducing the Entanglement Matrix, we introduce the concept of the 'degree of a midpoint'.

\subsection*{Degree of a Midpoint}

The degree of a midpoint is defined as the number of rays passing through a midpoint, M. So, suppose if a line is passing from a midpoint, P, then this is equivalent to 2 rays emerging from the midpoint, P. If two lines pass from a midpoint, Q, then this is equivalent to 4 rays emerging from that midpoint, Q and so on. In this way, we always have even degrees for a given midpoint.

Entanglement Matrix is a mathematical construct that gives information about all the possible entanglements present in given graph states of n - Qubits. Entanglement Matrix is constructed from graph states as follows:

\subsection*{The algorithm to Construct an Entanglement Matrix from the graph states:}
\begin{enumerate}
    \item  Consider a graph state of \( n \) qubits; we label the qubits in numerical order as 1, 2, 3, ... (either clockwise or anticlockwise) to get an ordered structure of qubits, and then take the midpoints of all the possible edges joining two qubits. So, for a graph state with n qubits (n - nodes), the maximum number of edges possible will be \( \binom{n}{2} \). So, there can be a maximum of \( \binom{n}{2} \) number of mid-points.\\
    We Classify the midpoints into two categories:
    \begin{itemize}
        \item \textbf{Primary Midpoints}: These midpoints are created from the edge joining the two adjacent nodes. More formally, these midpoints are created from the edges joining $k^{th}$ to $(k+1)^{th}$ nodes, \( \forall ~ k \in [1, n] \). The degree of primary midpoints will always be 2. Only Primary Midpoints are responsible for the bipartition of the graph state into two subsystems.
        \item \textbf{Secondary Midpoints}: These midpoints are created from the edge joining the two nodes that are not adjacent. More Formally, we can say that these midpoints are created from the edges joining $p^{th}$ to $q^{th}$, \( \forall \{p, q\} \in [1, n], q \notin \{p \pm 1\} \). The degree of secondary midpoints can be \( \geq 2 \). Secondary midpoints are responsible for the entanglement between the two non-adjacent nodes. They do not take part in the bipartition of a graph state
    \end{itemize}
\noindent 
Let us consider the two examples of graph states, with 5 qubits and 6 qubits, respectively:

\begin{figure}[h]
    \centering
    \includegraphics[width=0.47\textwidth]{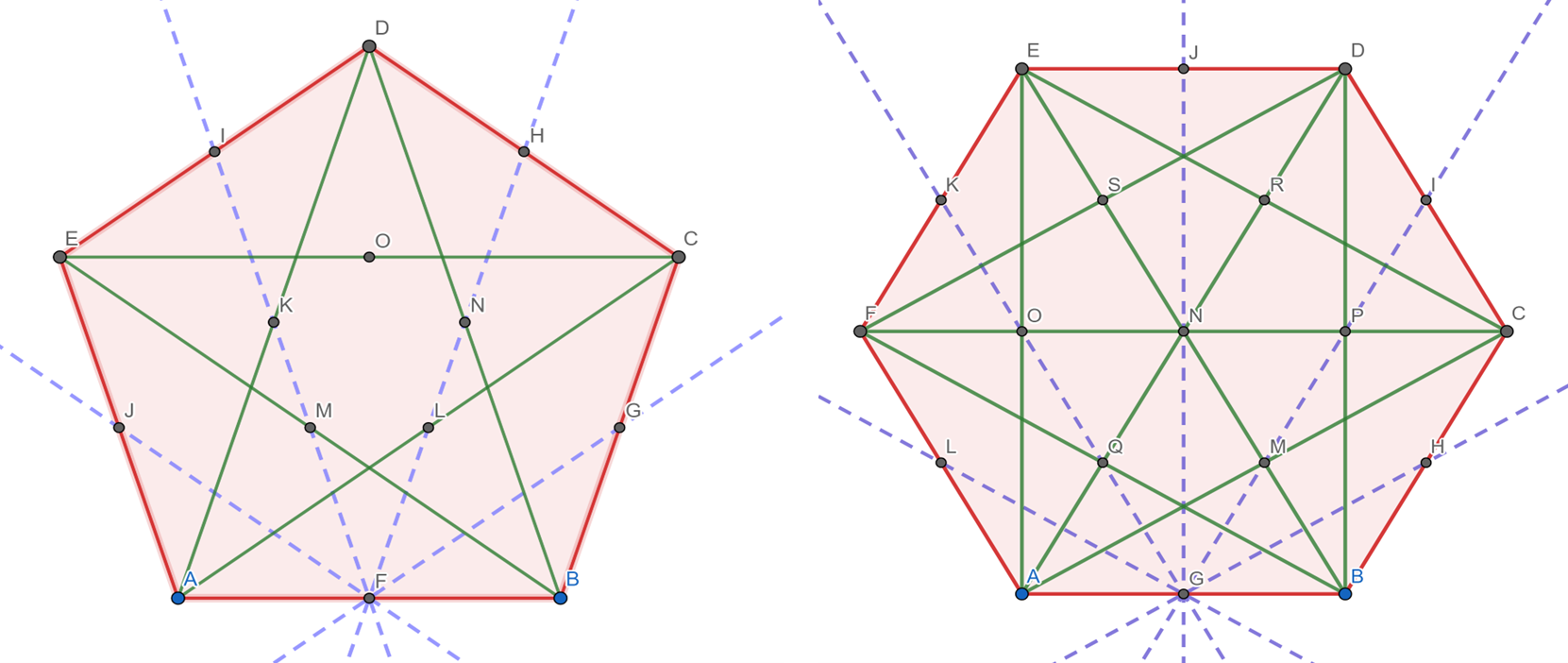}
    \caption{Graph States with 5 and 6 Qubits. The midpoints on the red line constitute the primary midpoints, whereas the midpoints inside the graph on the green line (joining two non-adjacent points) constitute the secondary midpoints. The purple dashed line shows all the possible bipartitions.}
\end{figure}

We first identify our graph state on the basis of the number of qubits (nodes) into odd graph or even graph with the following properties:
\begin{itemize}
    \item Odd Graph: In this graph, we have an odd number of qubits (nodes), and since we have an odd number of qubits, we do not have any edge passing through the centre of the graph, so all the midpoints will have degree 2 irrespective of their type as primary or secondary.
    \item Even Graph: In this graph, we will have an even number of Qubits (nodes), and so we have an edge connecting diametrically opposite Qubits (nodes) passing through the centre of the graph. So, for \( 2n \) qubits graph state, we will have the maximum degree of midpoint as \( 2n \), which is the centre of the graph state.
\end{itemize}
 \item Once we have identified the midpoints, beginning from primary midpoints, we start numbering them again, either clockwise or anticlockwise. This begins with \( 1' \), denoting the midpoint between qubits 1 and 2, and continues progressively as \( 2', 3', \ldots, n' \), where \( n \) represents the total number of qubits. Subsequently, the secondary midpoints, connecting non-adjacent qubits, are labelled analogously. Beginning with \( (n + 1)' \), denoting the midpoint between qubits 1 and 3, the labelling proceeds as \( (n + 2)', (n + 3)', \ldots \), until all secondary midpoints have distinct identifiers.
 \item We begin creating the entanglement matrix. Entanglement matrix is a \( n \times n \) square symmetric matrix, where \( n \) is the total number of midpoints in a given graph. Now, we join two primary midpoints (\( i' \) and \( j' \)), to divide the graph into two subsystems, such that each subsystem has same set of eigenvalues \( p_i \), from which we can calculate the entropy of formation used as a measure of entanglement. This entropy of formation can be Either Von-Neumann entropy or Rényi entropy varying over a parameter \( \alpha \). The result of this partition is added to the \textit{\{i'j'\}} and \textit{\{j'i'\}} entries of the entanglement matrix. In this way, we can find all the possible entanglements present in an arbitrary graph state.
\end{enumerate}

We can analyze the entries of the entanglement matrix of a given graph state as follows:
\begin{enumerate}
    \item The \{i'i'\} entry will give us the entanglement of the corresponding edge on which midpoint \( i' \) lies. For example, \{1'1'\} will give us the entanglement of the edge joining 1 and 2.
    \item The total entanglement of the system is given by the sum of all the elements on the primary diagonal and either of the upper or lower triangular parts.
    \item Primary Midpoints contribute to both diagonal and off-diagonal entries of the Entanglement Matrix. Whereas Secondary midpoints can only contribute to the diagonal entries.
\end{enumerate}

For secondary mid-midpoints possessing a degree (\( d \)), greater than 2, characterizing individual edge contributions to the overall entanglement becomes challenging. The in-distinguishability arises due to the superposition of multiple edges traversing the midpoint, effectively merging their entanglement signatures. To overcome this limitation, we propose taking the advantage of the adjacency matrix. By analyzing the matrix structure around the problematic midpoint, we can isolate the contributing edges and disentangle their individual entanglement characteristics.

\subsection*{Adjacency Matrix}

Let \( G \) be a graph with vertex set \( V = \{v_1, v_2, v_3, \ldots, v_n\} \). The adjacency matrix of \( G \) is the \( n \times n \) matrix whose \( (i, j) \) entry is:

\[
A(i, j) =
\begin{cases}
1 & \text{if } v_i \sim v_j \\
0 & \text{otherwise}
\end{cases}
\]

Since \( v_i \sim v_j \) if and only if \( v_j \sim v_i \), it follows that \( A(i, j) = A(j, i) \), and therefore \( A \) is a symmetric matrix, that is, \( A^T = A \). By definition, the indices of the non-zero entries of the \( i \)th row of \( A \) correspond to the neighbours of vertex \( v_i \). Similarly, the non-zero indices of the \( i \)th column of \( A \) are the neighbours of vertex \( v_i \).

Now, we use this definition the edges contributing to the overall entanglement at secondary midpoint of degree \( d (> 2) \) can be systematically identified and their individual characteristics of entanglement subsequently deconvoluted.

For example, for the given 4 qubit graph state, as shown in the Figure  ~\ref{fig:4qubiteg}, the Adjacency Matrix and the entanglement matrix can be written as:

\begin{figure}[h]
    \centering
    \includegraphics[width=0.25\textwidth]{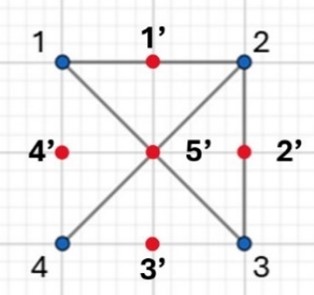}
    \caption{4 Qubit Graph State}
    \label{fig:4qubiteg}
\end{figure}
The adjacency matrix for the given graph state in Figure~\ref{fig:4qubiteg} is given as:
\[
\text{Adjacency Matrix } (A) = \begin{pmatrix}
0 & 1 & 1 & 0 \\
1 & 0 & 1 & 1 \\
1 & 1 & 0 & 0 \\
0 & 1 & 0 & 0
\end{pmatrix}
\]
Entanglement Matrix (E), for the above case can be written as:
\[
\text E = 
\begin{pmatrix}
\log 2 & \log 2 & \log 2 & \log 2 & 0 \\
\log 2 & \log 2 & \log 2 & \log 2 & 0 \\
\log 2 & \log 2 & 0 & 0 & 0 \\
\log 2 & \log 2 & 0 & 0 & 0 \\
0 & 0 & 0 & 0 & 2 \log 2
\end{pmatrix}
\]

From the above example, 1', 2', 3', 4', are the primary midpoints whereas, 5' is the secondary midpoint. So, from the entanglement matrix we found that the total entanglement of the given graph state (Figure  ~\ref{fig:4qubiteg}), is \( 9 \times \log 2 = 9 \). Notably, the 5'5' entry of the entanglement matrix gives us the total entanglement contribution from the edge CZ – 13 and CZ – 24 which is equal to \( 2 \log 2 \). But \( A_{24} \) (= \( A_{42} \)) and \( A_{13} \) (= \( A_{31} \)) entries of the adjacency matrix A will help us to distinguish between two different contributions to the entanglement at the secondary midpoint 5'. Another point to note is that since 5' is the secondary midpoint so, we only have the contribution to the diagonal entry 5'5', other entries corresponding to 5' are 0.

From the above example, we can conclude that for a given secondary midpoint M of degree \( d (>2) \) the total entanglement contributed by that midpoint to the entanglement matrix is equal to \( \frac{d}{2} \times \log 2 \). From the above example, we can also infer that wherever there is an edge connecting two qubits via a CZ gate, the von Neumann entanglement of that edge is \( \log_2 2 = 1 \).

Let us look at the below mentioned Tables ~\ref{3qclass} and  ~\ref{4qclass}, which gives us entanglement classification in three and four qubit non-isomorphic graph states based on the above developed formalism.

\begin{table*}[htb!]
    \centering
    \begin{tabular}{|>{\centering\arraybackslash}p{2.5cm}|>{\centering\arraybackslash}p{3cm}|>{\centering\arraybackslash}p{2.5cm}|>{\centering\arraybackslash}p{2.5cm}|>{\centering\arraybackslash}p{3cm}|}
        \hline
        \textbf{Non-Isomorphic Class} & \textbf{Total Number of Canonical States} & \textbf{Total Entanglement} & \textbf{Entanglement Class} & \textbf{Description} \\
        \hline
        Class 1 & 1 & 0 & Class 1 & Fully Separable\\
        Class 2 & 3 & $3 \log_2 {2}$ & Class 2 & Bi-Separable\\
        Class 3 & 3 & $5 \log_2 {2}$ & Class 3 & Entangled\\
        Class 4 & 1 & $6 \log_2 {2}$ & Class 4 & Fully Entangled\\
        \hline
    \end{tabular}
    \caption{Entanglement Classification in 3 Qubit Non-Isomorphic Graph States}
    \label{3qclass}
\end{table*}

\begin{table*}[htb!]
  \centering
  \includegraphics[width=0.75\textwidth]{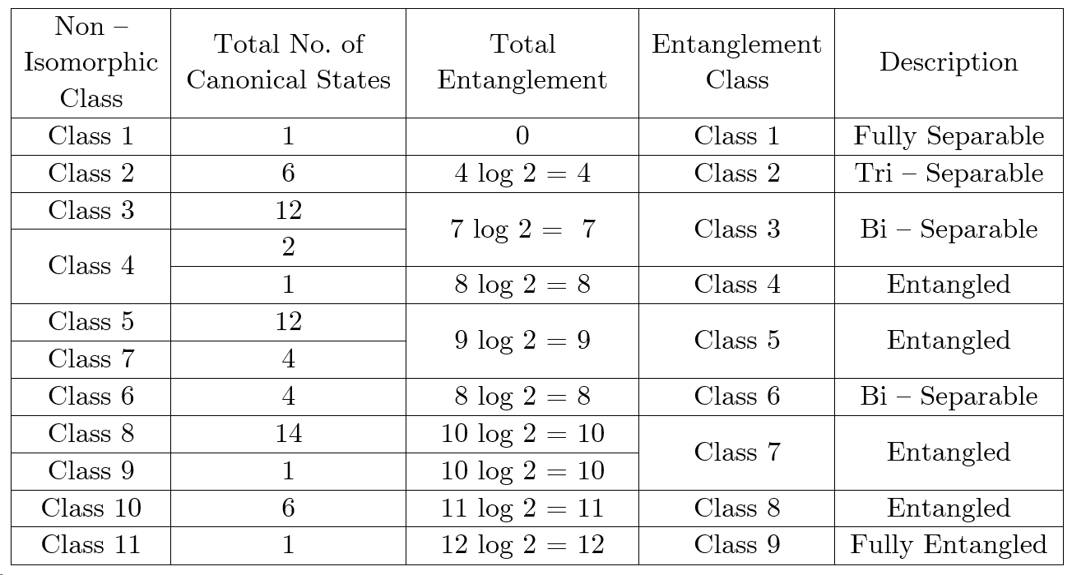}
  \caption{Entanglement Classification in 4 Qubit Non‑Isomorphic Graph States}
  \label{4qclass}
\end{table*}

Now, we will focus on maximum possible entanglement in n-qubit graph state:
\section{Maximum Entanglement}

For all Graph States, the maximum entanglement occurs when every node (qubit) is connected to every other node (qubit) in the graph state via a CZ Gate. Now, for a given graph state with N qubits (nodes), the maximum number of edges that connects one qubit to another via a CZ Unitary gate will be \( \binom{N}{2} \). So, the graph with maximum number of edges contains the maximum entanglement and total edges in such graph will be \( \binom{N}{2} \).

\begin{figure*}[htb]
  \centering
  \includegraphics[width=\textwidth]{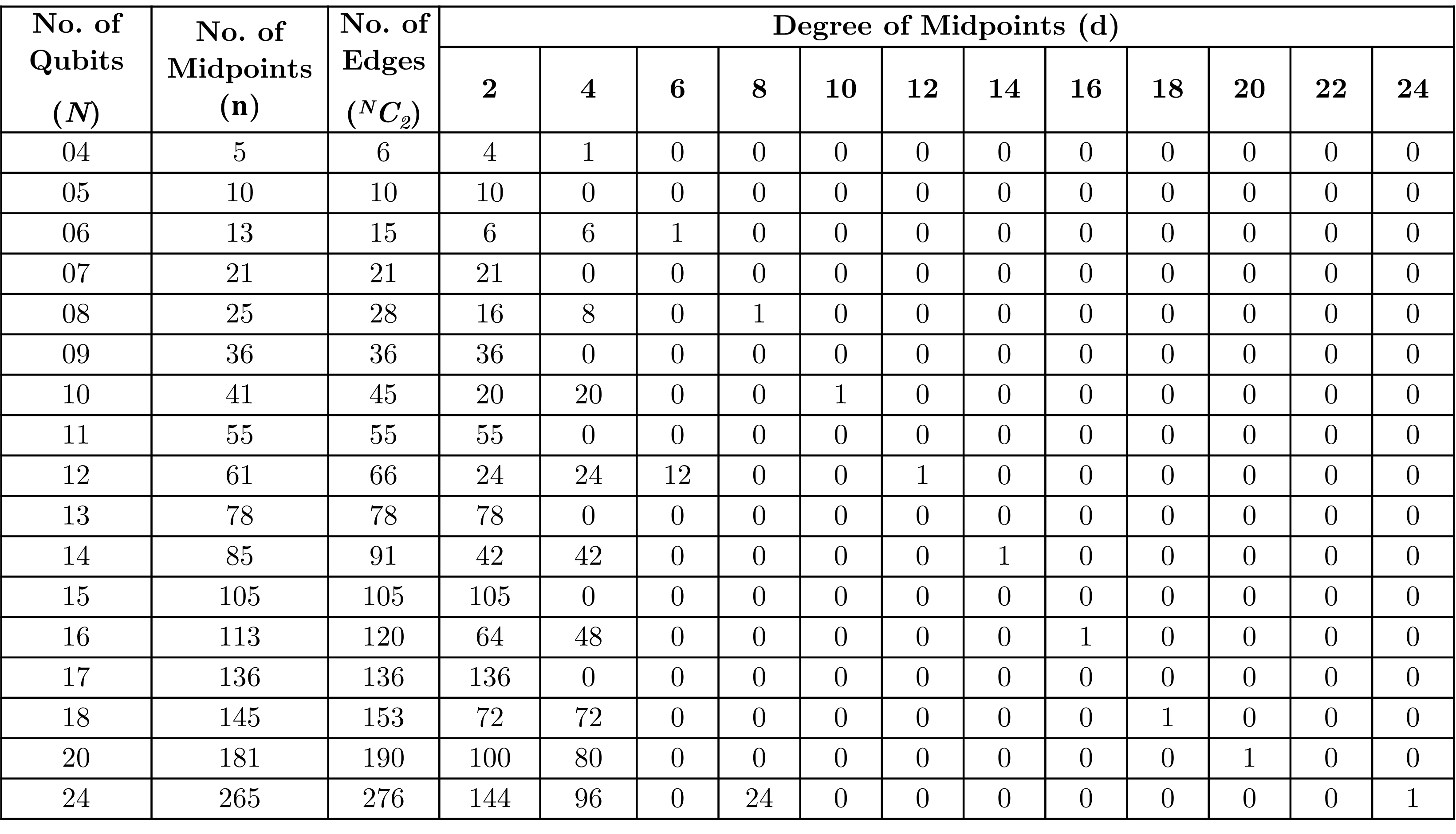}
  \caption{Total midpoints and number of midpoints of degree $d$, for the initial $n$‑qubit ($n = 24$) graph states.}
  \label{fig:midpd}
\end{figure*}

To begin with, consider the following Table~\ref{fig:midpd}, which represents the total midpoints and number of midpoints of degree \( d \), for initial n-qubit graph states.

We will continue our analysis for the maximum entanglement in a given graph state into two different categories, one belonging to Odd Graph State and the other belonging to 
Even Graph State.

\subsection*{Maximum Entanglement in Odd Graph States}

In case of Odd Graph State of \( N \) qubits, the maximum number of edges will be \( \binom{N}{2} \). Since in Odd Graph State, all the midpoints that occur will be of degree 2. Since, each edge will contribute to a single midpoint, so, there will be \( \binom{N}{2} \) midpoints. The size of entanglement matrix will be equal to the number of midpoints in the given graph. Since, the total number of midpoints in a graph of \( N \) qubits will be \( \binom{N}{2} \), so the size of entanglement matrix will be \( \binom{N}{2} \times \binom{N}{2} \). In order to determine the Maximum entanglement captured by a Odd graph state, we will consider two cases, one in which belongs to entanglement contribution by the primary midpoints and the second which belongs to the secondary midpoints.
\begin{enumerate}
    \item \textbf{Case 1:} Entanglement captured by the primary midpoints:\\
    For a graph state of \( N \) qubits, the first \( N \times N \) block belongs to the primary midpoints. So, all the entries of this submatrix of the entanglement matrix of any graph of \( N \) Qubits (\( N = \) Odd) will be \( \log 2 \) (Considering von-Neumann Entropy).

    Since, the total entanglement of any given graph state is calculated from its entanglement matrix by taking the sum of diagonal elements and either of upper or lower triangular matrix elements (as the entanglement matrix is a symmetric matrix).

    We know that the number of triangular matrix entries in an \( N \times N \) matrix. So, without the diagonal it is \( \frac{N(N-1)}{2} \) and including the diagonal it is \( \frac{N(N+1)}{2} \).

    So, for primary midpoints, the first \( N \times N \) block will be populated by its corresponding Von-Neumann Entropy value which is \( \log 2 \). Which implies that Total entanglement captured by primary midpoints will be: \( \frac{N(N+1)}{2} \times \log 2 \).

    \item \textbf{Case 2:} Entanglement captured by Secondary midpoints:

    These midpoints do not contribute to the off-diagonal elements of entanglement matrix and so, the entanglement captured by secondary midpoints will belong to diagonal entries of the entanglement matrix.

    We know the size of entanglement matrix is \( \binom{N}{2} \times \binom{N}{2} \) and out of which first submatrix of \( N \times N \) entries will be contributed by primary midpoints and the rest belongs to secondary midpoints. Since, secondary midpoints can only contribute to the diagonal elements, so, the number of off diagonal entries for secondary midpoints will be \( \binom{N}{2} - N \).

    The total entanglement captured by primary midpoints will be:

    \begin{align}
    E_{Max} &= \frac{N(N+1)}{2} \times \log_2 2 + \left[ \frac{N(N-1)}{2} - N \right] \times \log_2 2 \notag\\
    &= N^2 - N
    \end{align}

    Hence,

    The maximum entanglement captured in case of \( N \) qubit Odd graph state is \( N^2 - N \).
\end{enumerate}

Plotting \( E_{Max} \) as a function of \( N \), we get:

\begin{figure}[h]
    \centering
    \includegraphics[width=0.5\textwidth]{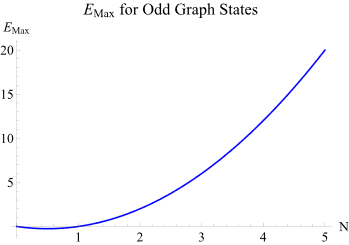}
    \caption{Maximum Entanglement in case of Odd Graph States. Following the quadratic relation in N (Number of Qubits), we have two zeroes, one is at \( N = 0 \) and other one is at \( N = 1 \), verifying that a single qubit cannot be entangled.}
\end{figure}

Now, we will proceed onto our analysis for the maximum entanglement in a given graph state with even number of qubits. 

\subsection*{Maximum Entanglement in Even Graph States}

If we calculate the maximum entanglement based on formula we derived for odd graph states, Equation (1), then we undercalculate the maximum entanglement. This happens because, the midpoints, specifically secondary midpoints, are not of degree 2 anymore, the degree of secondary midpoints in case of Even graph states is more than or equal to 2 (\( d \geq 2 \)). So, midpoints of degree \( d > 2 \) are responsible for increasing the total maximum entanglement in Even Graph states.

Before proceeding we will have a look at some of the Maximally entangled graph states with even number of qubits.

\begin{figure}[h]
    \centering
    \includegraphics[width=0.47\textwidth]{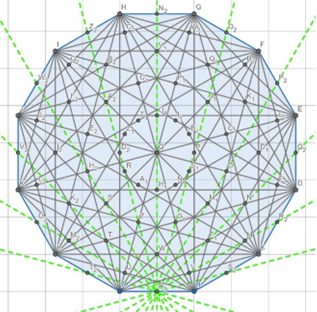}
    \caption{Example of a graph state with even number of Qubits (12 Qubits). The green dashed line represents possible bipartitions of graph state.}
\end{figure}

Notice that the midpoints of all the edges occur/ form in concentric circles around the midpoint, of which the outermost circle contains the primary set of midpoints, and rest of the concentric circles including the middle centre point belong to secondary set of midpoints. Before deriving general formula, let us look at some of the peculiar observations.

Notice that, the midpoints on a particular concentric circles are exactly same in number and equal to the number of qubits we have considered. Now, if we start noting down the degree of midpoints, as we go from the outermost circle (primary midpoints) to the innermost central point then we observe some pattern in the degree of midpoints encountered as we go in to the central point.

Interestingly, the concentric circles share an intriguing property: the number of midpoints on each circle precisely matches the number of qubits considered. Furthermore, analyzing the degrees of these midpoints as we progress from the outermost "primary midpoints" towards the central point reveals a distinct yet 'peculiar' pattern.

The following Table~\ref{fig:degeven} provides some insight into first few even qubits and their corresponding degree of midpoints encountered as we go from outermost to the central point.

\begin{table*}[htb!]
    \centering
    \includegraphics[width=1\textwidth]{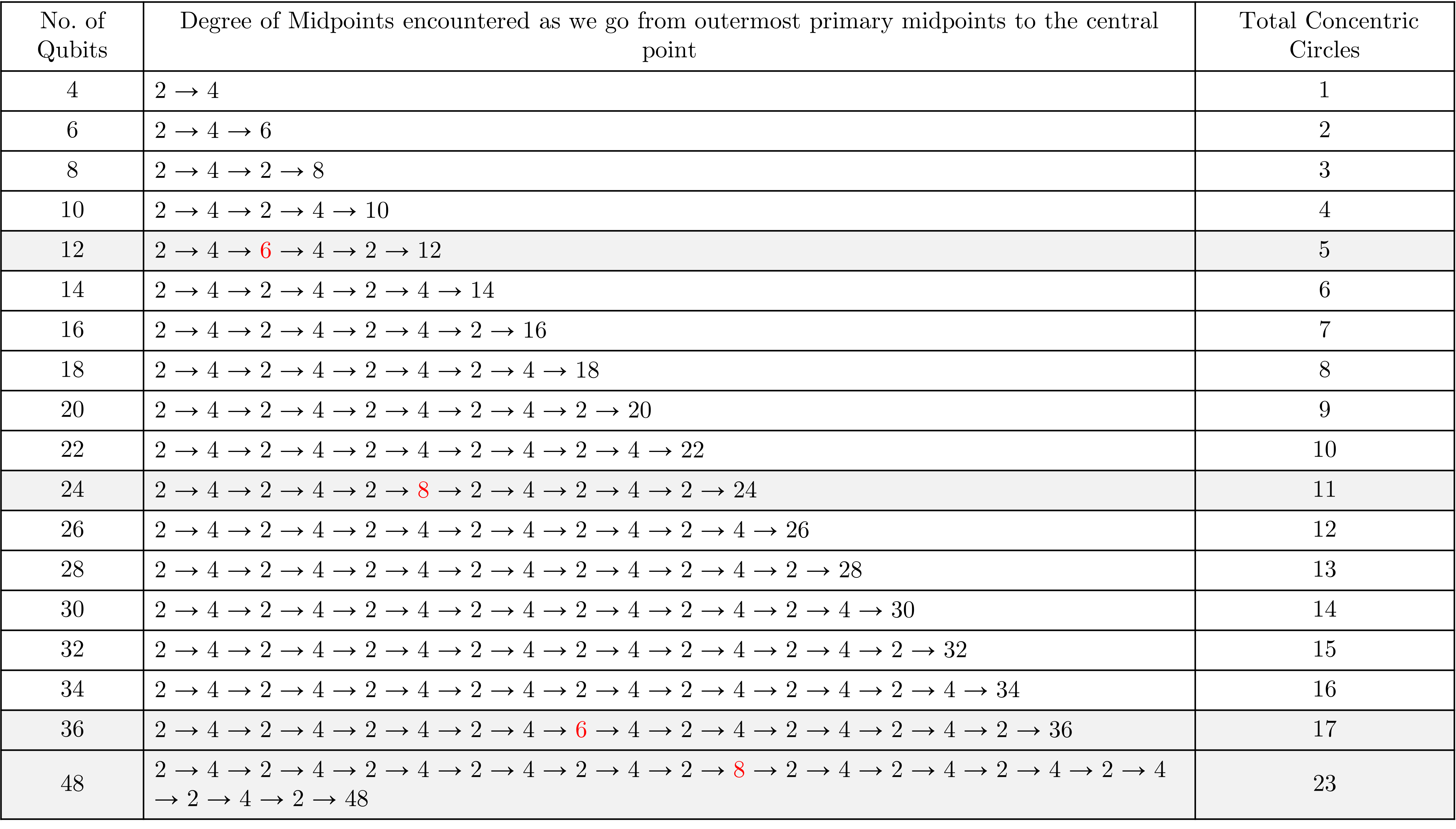}
    \caption{Observation table for graph states with even number of qubits.  We start with outermost primary midpoint with a degree of 2 and progresses inwards.  The sequence follows an alternating pattern of midpoints having degrees 2 and 4.  Importantly, the degree of the last midpoint (central point) is equal to with the total number of qubits in the graph state.  The third column specifies the number of concentric circles surrounding the central midpoint on which these midpoints reside.}
    \label{fig:degeven}
\end{table*}

\subsubsection*{Total number of midpoints:}

Let us find out the total number of midpoints for a given even graph state (N qubits) as a function of N as:

For Even graph state with N qubits, we have, \( \left(\frac{N}{2} - 1\right) \) concentric circles. On each concentric circle we have N midpoints with degree \( d \geq 2 \).

Since we also have a 'central midpoint' of degree N, total number of Midpoints for a given graph state of N qubits will be: \( M = N \left(\frac{N}{2} - 1\right) + 1 \). Hence our entanglement matrix will be of size \( M \times M \), where \( M = N \left(\frac{N}{2} - 1\right) + 1 \).

We will take the graph states in which total qubits are integral multiple of 12 as a special case later. For now, we will focus on the graph states in which total qubits are not the integral multiple of 12.

\subsubsection*{Multiplicity of a midpoint of degree 'd'.}

Let us define the multiplicity (\( m \)) of degree (\( d \)): Multiplicity (\( m \)) of degree (\( d \)) is defined as \( m = \frac{d}{2} \), which shows that if the degree of a midpoint is \( d \), then there are \( m = \frac{d}{2} \) lines passing through that midpoint, and each line is contributing to the entanglement.

So, for a secondary midpoint, 'S' of degree' \( d \) and multiplicity \( m = \frac{d}{2} \), the ‘SS’ entry (diagonal entry), of the entanglement matrix will be \( m \cdot \log_2 2 \) (Considering Von-Neumann Entropy).

\subsubsection*{Bipartite Entanglement Calculation}

Now, from Table  ~\ref{fig:degeven}, we observe that, midpoints of degree 2 and 4 occur alternatively on the concentric circles beginning from the outermost primary midpoint of degree 2. We know that for even graph state with N qubits, we have, \( \left(\frac{N}{2} - 1\right) \) concentric circles. So, let us denote total concentric circles as \( C = \left(\frac{N}{2} - 1\right) \).

\begin{enumerate}
    \item \textbf{Case 1}: If \( \frac{C}{2} \) is integer:
        \begin{itemize}
        \item Total number of midpoints of degree 2 = \( N \cdot \frac{C}{2} \)
        \item Total number of midpoints of degree 4 = \( N \cdot \frac{C}{2} \)
        \end{itemize}
    So, we have equal number of midpoints of degree 2 and 4 in this case. We also have a central midpoint of degree N.
    \item \textbf{Case 2}: If \( \frac{C}{2} \) is not an integer:
        \begin{itemize}
            \item Total number of midpoints of degree 2 = \( N \cdot \left\lceil \frac{C}{2} \right\rceil = N \cdot \text{ceil} \left(\frac{C}{2}\right) \)
            \item Total number of midpoints of degree 4 = \( N \cdot \left\lfloor \frac{C}{2} \right\rfloor = N \cdot \text{floor} \left(\frac{C}{2}\right) \)
        \end{itemize}
    Where, \( \left\lceil x \right\rceil = \text{ceil}(x) \) and \( \left\lfloor x \right\rfloor = \text{floor}(x) \) are the ceiling function and floor function of \( x \) respectively. As we know from the properties of Ceiling and Floor Functions, that, \( \left\lceil x \right\rceil = \left\lfloor x \right\rfloor = x \) if \( x \) is an integer.    
\end{enumerate}

So, the above two cases can be written in combined form as:\\
For a given \( C = \left(\frac{N}{2} - 1\right) \); where \( C \) denotes the total number of concentric circles,

\begin{itemize}
    \item Total number of midpoints of degree 2 = \( N \cdot \left\lceil \frac{C}{2} \right\rceil = N \cdot \text{ceil} \left(\frac{C}{2}\right) \)
    \item Total number of midpoints of degree 4 = \( N \cdot \left\lfloor \frac{C}{2} \right\rfloor = N \cdot \text{floor} \left(\frac{C}{2}\right) \)
\end{itemize}

Where, multiplicity of midpoints of degree 4 is 2 and multiplicity of midpoints of degree 2 is 1.\\

Now, let us calculate the maximum possible entanglement for a given graph state of even qubits:\\

For entanglement matrix, since only primary midpoints are responsible for bipartition and hence contribute to the off-diagonal entries of the entanglement matrix. Whereas secondary midpoints contribute only to the diagonal entries of entanglement matrix. So, consider the following:

For a given graph state of \( N \) Qubits (\( N = \) even), having ‘\( C \)’ concentric circles,
Total number of primary midpoints = \( N \)

Total entanglement contribution from primary midpoints: \( \frac{N(N+1)}{2} \times \log_2 2 \)

Total number of Secondary midpoints: \( (N \cdot \left\lceil \frac{C}{2} \right\rceil - N) + (N \cdot \left\lfloor \frac{C}{2} \right\rfloor) \)

Entanglement Contribution from Secondary Midpoints (Only diagonal Entries):

\( (N \cdot \left\lceil \frac{C}{2} \right\rceil - N) \times \log_2 2 + (N \cdot \left\lfloor \frac{C}{2} \right\rfloor) \times 2 \times \log_2 2 \)

Now, since we have ceiling and floor functions involved in our calculations, we will take two cases:
\begin{enumerate}
    \item \textbf{Case 1:} When \( \frac{C}{2} \) is an integer:\\
    In this case, \( \left\lceil \frac{C}{2} \right\rceil = \left\lfloor \frac{C}{2} \right\rfloor = \left(\frac{C}{2}\right) \):
    Entanglement from Secondary midpoints:
    \begin{align}
    &(N \cdot \left\lceil \frac{C}{2} \right\rceil - N) \times \log_2 2 + (N \cdot \left\lfloor \frac{C}{2} \right\rfloor) \times 2 \times \log_2 2 \notag\\
    &= (N \cdot \left(\frac{C}{2}\right) - N) + 2 \cdot (N \cdot \left(\frac{C}{2}\right)) \notag\\
    &= \left(\frac{N}{2}\left(\frac{N}{2} - 1\right) - N\right) + 2 \left(\frac{N}{2}\left(\frac{N}{2} - 1\right)\right) \notag\\
    &= \left(\frac{N^2}{4} - \frac{N}{2} - N\right) + \left(\frac{2N^2}{4} - \frac{2N}{2}\right) \notag\\
    &= \frac{3N^2}{4} - \frac{5N}{2}
    \end{align}
    Total Maximum Entanglement:
    \begin{align}
    E_{max} &= \frac{N^2}{2} + \frac{N}{2} + \frac{3N^2}{4} - \frac{5N}{2} \notag\\
    &= \frac{5N^2}{4} - \frac{4N}{2}
    \end{align}
    \item \textbf{Case 2:} When \( \frac{C}{2} \) is NOT an integer:\\
    In this case, entanglement from the Secondary midpoints is given by:
    \begin{align}
    &(N \cdot \left\lceil \frac{C}{2} \right\rceil - N) \times \log_2 2 + (N \cdot \left\lfloor \frac{C}{2} \right\rfloor) \times 2 \times \log_2 2 \notag\\
    &= (N \cdot \left(\frac{C+1}{2}\right) - N) + 2 \cdot (N \cdot \left(\frac{C-1}{2}\right)) \notag\\
    &= \left(\frac{N}{2} \times \left(\frac{N}{2} - 1\right) - N\right) + 2 \left(\frac{N}{2} \left(\frac{N}{2} - 2\right)\right) \notag\\
    &= \frac{3N^2}{4} - 3N
    \end{align}
    Total Maximum Entanglement:
    \begin{align}
    E_{max} &= \frac{N^2}{2} + \frac{N}{2} + \frac{3N^2}{4} - 2N \notag\\
    &= \frac{5N^2}{4} - \frac{5N}{2}
    \end{align}
\end{enumerate}

Finally, adding the entanglement from the ‘central midpoint’, the total maximum entanglement for a given graph state with even number of qubits (\( N \)) is given by:

\[
E_{max} = 
\begin{cases}
\frac{5N^2}{4} - 2N + \frac{N}{2} = \frac{5N^2}{4} - \frac{3N}{2} & \text{for } \frac{C}{2} \in \mathbb{Z}^+ \\
\frac{5N^2}{4} - \frac{5N}{2} + \frac{N}{2} = \frac{5N^2}{4} - \frac{4N}{2} & \text{for } \frac{C}{2} \notin \mathbb{Z}^+
\end{cases}
\]

Now, if we plot \( E_{max} \) as a function of \( N \):

\begin{figure}[h]
    \centering
    \includegraphics[width=0.5\textwidth]{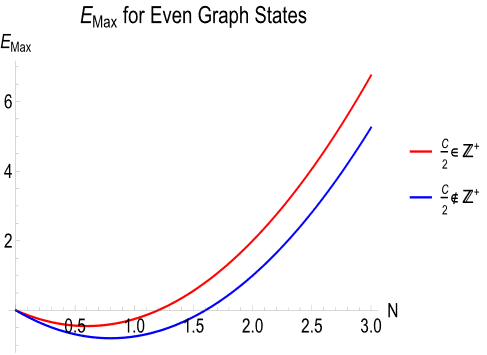}
    \caption{Maximum Entanglement in case of Even Graph States.  Following the Quadratic Relation in N (Number of Qubits).  Entanglement is more when C/2 is an integer compared to when C/2 is not an integer.}
\end{figure}

\subsection*{A Special case of Even graph States:}
Now, we will see that for some special cases. In which the total number of qubits are an integral multiple of 12, we get a midpoint in between whose degree is neither 2 nor 4, but either 6 or 8 depending on whether that midpoint is an even or odd multiple of 12 respectively.

A more detailed observation from the Table  ~\ref{fig:degeven} reveals that whenever the total number of qubits is even multiple of 12 then we have a midpoint of degree 4 replaced by a midpoint of degree 8, whereas whenever the total number of qubits is odd multiple of 12 then we have a midpoint of degree 2 replaced by a midpoint of degree 6. Only a single midpoint in a midway between is replaced as mentioned above. The reason why this happens is still not clear.

So, following the above considerations, we modify our formulation for these special cases. Consider the following Table~\ref{12mul}.

\begin{table}[h]
    \centering
    \begin{tabular}{|>{\centering\arraybackslash}p{1.4cm}|>{\centering\arraybackslash}p{2.4cm}|>{\centering\arraybackslash}p{1.0cm}|>{\centering\arraybackslash}p{2.5cm}|}
        \hline
        \textbf{No. of Qubits} & \textbf{No. of Concentric Circles (C)} & \textbf{C/2} & \textbf{Midpoint Replacement} \\
        \hline
        12 & 5 & 2.5 & 2 replaced by 6 \\
        24 & 11 & 5.5 & 4 replaced by 8\\
        36 & 17 & 8.5 & 2 replaced by 6\\
        48 & 23 & 11.5 & 4 replaced by 8\\
        \hline
    \end{tabular}
    \caption{Observation table shows the number of concentric circles and the midpoint replacement depending on whether total qubits are even or odd multiple of 12.}
    \label{12mul}
\end{table}

Now, from Table~\ref{12mul}, we see that \( \frac{C}{2} \notin \mathbb{Z}^+ \), therefore, the second equation, \( E_{max} = \frac{5N^2}{4} - 2N \), needs to be modified. Let us consider two cases:

\textbf{CASE 1:} When the number of qubits is even multiple of 12.

When the total number of qubits is even multiple of 12 then we have a midpoint of degree 4 replaced by a midpoint of degree 8. This gives us:

\begin{align}
  &\begin{aligned}[b]
    \bigl(N\lceil C/2\rceil - N\bigr)\,\log_{2}2
    + \bigl(N\lfloor C/2\rfloor - N\bigr)\,2\,\log_{2}2\\
    + 4\,N\,\log_{2}2
  \end{aligned}\notag\\
  &=\;\begin{aligned}[b]
    \bigl(N\cdot\tfrac{C+1}{2}-N\bigr)
    +2\bigl(N\cdot\tfrac{C-1}{2}\bigr)
    -2N+4N
  \end{aligned}\notag\\
  &=\;\begin{aligned}[b]
    \bigl(\tfrac{N}{2}(\tfrac{N}{2}-1)-N\bigr)
    +\bigl(\tfrac{N}{2}(\tfrac{N}{2}-2)\bigr)\,2
    +2N
  \end{aligned}\notag\\
  &=\;\frac{3N^{2}}{4}-N
\end{align}

Total Maximum Entanglement:

\begin{align}
    E_{max} &= \frac{N^2}{2} + \frac{N}{2} + \frac{3N^2}{4} - 2N^2 + \frac{N}{2} \notag\\
    &= \frac{5N^2}{4}
\end{align}

\textbf{CASE 2:} When the number of qubits is odd multiple of 12.

When the total number of qubits is odd multiple of 12 then we have a midpoint of degree 2 replaced by a midpoint of degree 6. This gives us:

\begin{align}
  &\begin{aligned}[b]
    \bigl(N\lceil C/2\rceil - 2N\bigr)\log_{2}2
    + 3N\,\log_{2}2
    + \bigl(N\lfloor C/2\rfloor\bigr)\,2\,\log_{2}2
  \end{aligned} \notag\\
  &=\;\begin{aligned}[b]
    \bigl(N\cdot\tfrac{C+1}{2} - N\bigr) - N + 3N
    + 2\cdot\bigl(N\cdot\tfrac{C-1}{2}\bigr)
  \end{aligned} \notag\\
  &=\;\begin{aligned}[b]
    \bigl(\tfrac{N}{2}(\tfrac{N}{2}-1) - N\bigr)
    + \bigl(\tfrac{N}{2}(\tfrac{N}{2}-2)\bigr)\cdot 2 + 2N
  \end{aligned} \notag\\
  &=\;\frac{3N^2}{4} - N
\end{align}

Total Maximum Entanglement:

\begin{align}
    E_{max} &= \frac{N^2}{2} + \frac{N}{2} + \frac{3N^2}{4} - 2N^2 + \frac{N}{2} \notag\\
    &= \frac{5N^2}{4}
\end{align}

Hence, we get maximum entanglement as \( E_{max} = \frac{5N^2}{4} \), irrespective of whether the total number of qubits are even or odd multiples of 12! Even though in both the cases we are replacing the midpoints of different degrees (2 and 4) with midpoints of some other degrees (6 and 8), the overall entanglement stays the same!

\section{Results and Discussion}

\begin{figure}[h]
    \centering
    \includegraphics[width=0.5\textwidth]{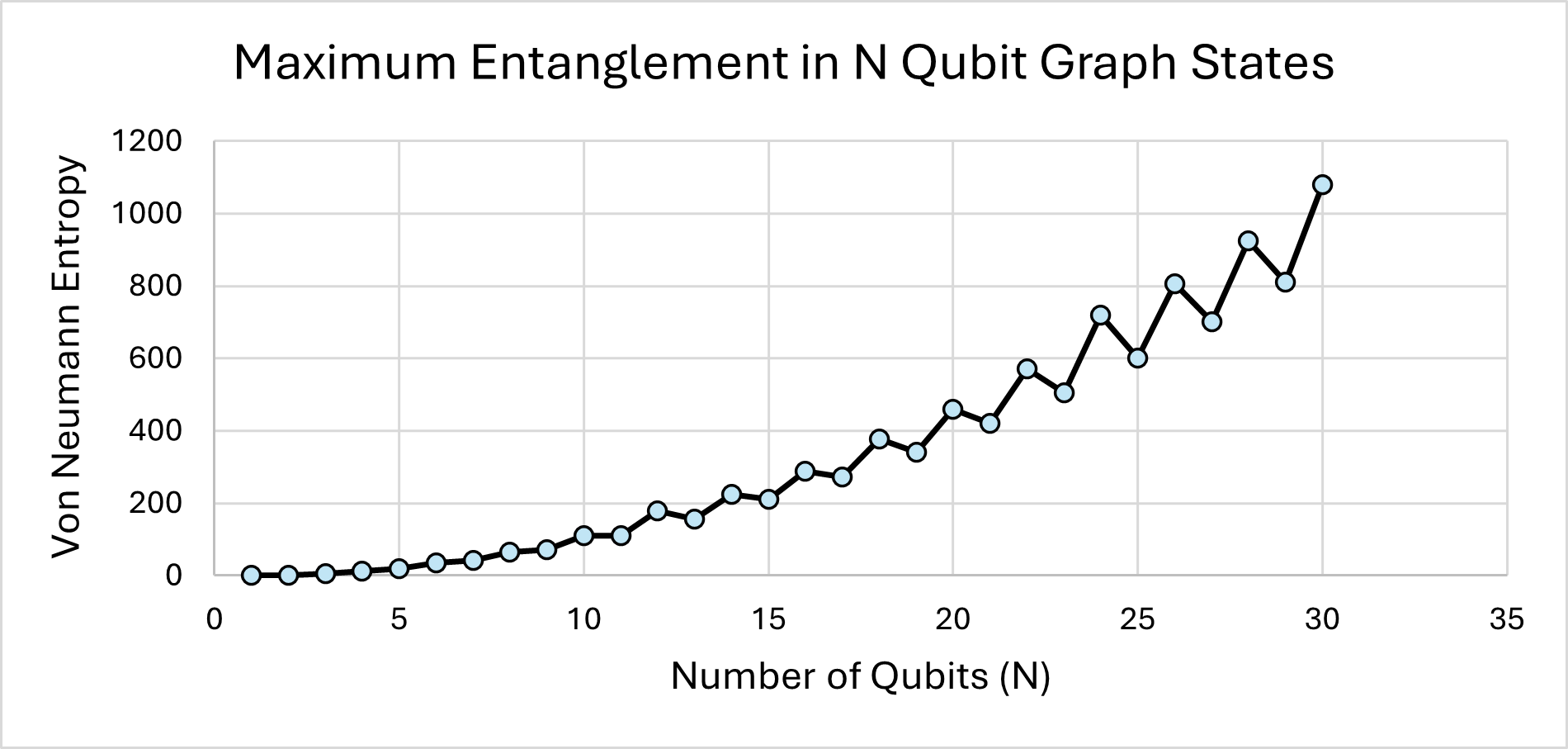}
    \caption{Maximum Entanglement as a function of N qubits. For Graph states, with odd number of qubits, the entanglement decreases compared to the even qubit graph states.}
    \label{fig:emt}
\end{figure}

In this paper, we quantified \( N \) qubit graph state using Von – Neumann entropy of formation. Our computations reveals that the entanglement follows different behaviour depending on whether \( N \) is even or odd. The Maximum entanglement varies quadratically with \( N \) (Number of Qubits). The anomaly arises when the total number of qubits are an integral multiple of 12, in that case, we see that entanglement increases. Figure  ~\ref{fig:emt} provide a distinction between the entanglement pattern we observed as a function of \( N \).

The combined result of our analysis is shown below:

\[
E_{max} =
\begin{cases}
N^2 - N & \text{for } N = \text{Odd} \\
\frac{5N^2}{4} - \frac{3N}{2} & \text{for } N = \text{Even, } \frac{C}{2} \in \mathbb{Z}^+ \\
\frac{5N^2}{4} - 2N & \text{for } N = \text{Even, } \frac{C}{2} \notin \mathbb{Z}^+ \\
\frac{5N^2}{4} & \text{for } N = \text{Even, multiple of 12}
\end{cases}
\]

If we plot \( E_{max} \) as a function of \( N \), then we get the following graph:

\begin{figure}[h]
    \centering
    \includegraphics[width=0.5\textwidth]{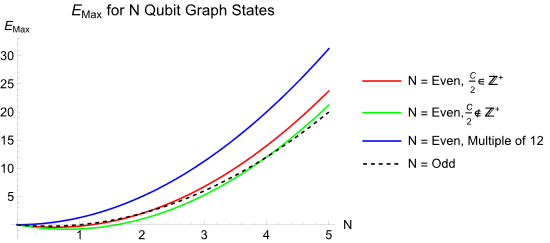}
    \caption{Maximum Entanglement captured in \( N \) Qubit Graph States}
\end{figure}

Now,  Figure~\ref{fig:emt}, represents maximum entanglement as a function of number of qubits. Which means that for a given graph state, the entanglement captured must be less than or equal to the maximum entanglement. However, when we convolute all the four graphs and form a single graph, we get a clearer behaviour of entanglement of N qubits graph state.

\begin{figure}[h]
    \centering
    \includegraphics[width=0.5\textwidth]{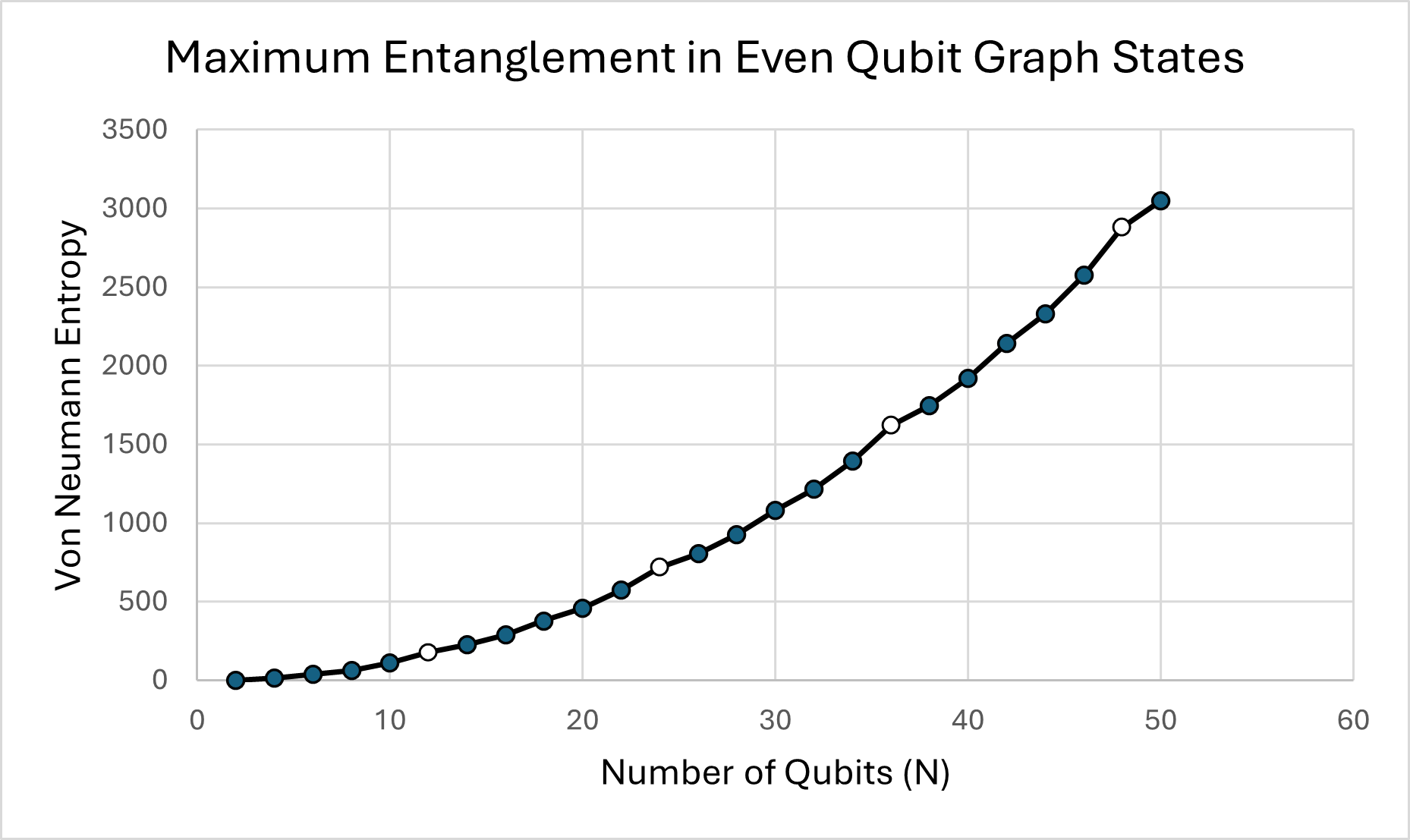}
    \caption{Maximum Entanglement as a function of N even Qubits.  Notice that, entanglement is slightly higher  for the Graph states in which qubits are integral multiple of 12.}
    \label{fig:eventot}
\end{figure}

However, to see the variation in even qubits graph states, we plot the entanglement measure against the even qubits graph states as shown in Figure~\ref{fig:eventot}.

\FloatBarrier

\section{Conclusion}
Our analysis demonstrates that the entanglement in graph states exhibits a quadratic dependence on the number of qubits \(N\). Specifically, graph states with an odd number of qubits display a reduction in the overall maximum entanglement compared to their even-numbered counterparts. This reduction arises because, in odd-qubit graph states, all midpoints possess degree 2, thereby limiting the entanglement contribution from each midpoint. In contrast, even-qubit graph states feature midpoints with degrees greater than or equal to 2, which enhances the overall entanglement in the system due to increased connectivity.

A noteworthy observation is that graph states with a total number of qubits that are integer multiples of 12 exhibit a unique structural feature: the presence of a midpoint with degree 6 or 8 at intermediate positions, depending on whether the multiple of 12 is odd or even, respectively. The underlying reason for this peculiar behavior remains an open question.

Furthermore, we conclude that graph states with the total number of qubits being a multiple of 12 attain slightly higher entanglement than other even graph states. This enhancement is attributed to the occurrence of midpoints with degree 6 or 8, as detailed in Table~\ref{12mul}, which depends on whether the total number of qubits is an odd or even multiple of 12. All other, less entangled states fall below these maximum entanglement values.

Thus, our analysis identifies the ``maximum entangled class" of \(N\)-qubit graph states and establishes a rigorous formalism for quantifying and classifying entanglement in such systems using the Entanglement Matrix approach.

Finally, the recent experimental realization of graph‐state entanglement in atomic ensembles demonstrates a clear route to implement our entanglement‐matrix framework on physical platforms \cite{Perlin2024Experimental}. By combining photon‐mediated spin–nematic squeezing with programmable spin rotations, one can imprint any desired adjacency‐matrix eigenmode and verify genuine multipartite entanglement via nullifier variances and EPR‐steering criteria. This synergy between theory and cavity‐QED techniques not only validates our classification scheme but also establishes a scalable pathway for benchmarking resource states in continuous‐variable quantum computing and sensing.

\bibliographystyle{apsrev4-1}
\bibliography{ref}
\end{document}